\RequirePackage[2020-02-02]{latexrelease}

\documentclass[usenames,dvipsnames, aps, pra, preprint, groupedaddress, amsfonts,
               amsmath, amssymb, showpacs, nofootinbib]{revtex4-1}
\usepackage{microtype}
\usepackage{lmodern}
\usepackage[T1]{fontenc}
\usepackage{graphicx}
\usepackage[margin=0.9in]{geometry}

\usepackage{epsfig}

\usepackage{siunitx}
\usepackage{epstopdf}
\usepackage{changepage}
\usepackage[utf8]{inputenc}
\usepackage[T1]{fontenc}
\usepackage{xcolor}
\usepackage{hyperref}

\usepackage{color}% \textcolor{#1}{#2}

\begin{document}
\raggedbottom
\title{Partial-wave approach to the Stark resonance problem of the water molecule}

\author{Patrik Pirkola}
\email[]{patpirko@my.yorku.ca}
\author{Marko Horbatsch}
\email[]{marko@yorku.ca}
\affiliation{Department of Physics and Astronomy, York University, Toronto, Ontario M3J 1P3, Canada}
\date{\today}
\begin{abstract}
A partial-wave method is developed to deal with small molecules dominated by a central atom as an extension of earlier single-center methods. In particular, a model potential for the water molecule is expanded over a basis of spherical harmonics.
A finite element method is employed to generate local polynomial
functions in subintervals to represent the radial part of the wavefunction. The angular parts of the wavefunction are represented by spherical harmonics. The problem of Stark resonances is treated with the exterior complex scaling method which incorporates a wavefunction discontinuity at the scaling radius.
The resultant non-hermitian matrix eigenvalue problem yields resonance
positions and widths (decay rates).
We present these DC Stark shifts and exponential decay rates for the valence
orbitals $1b_1$, $3a_1$, and the bonding orbital $1b_2$.
Furthermore, comparison is made with total molecular decay rates 
and DC shifts obtained recently within the Hartree-Fock and coupled-cluster approaches. 
\end{abstract}
%
%\pacs{34.10.+x, 34.50.Gb, 34.70.+e, 36.40.-c}
%

\maketitle
\section{Introduction}
\label{intro}
The study of molecules in external fields, and particularly with lasers and X rays, but also in collisions with charged particles has received
considerable attention in recent years.
The investigation of water molecules is
motivated by its relevance for radiation medicine,
e.g., inner-shell electron removal by hard radiation
has been reported~\cite{PhysRevX.11.041044}. Collisions of electrons (and
positrons) with water molecules are also an active area of
current research~\cite{ElectronsWater2021}.

The problem of multi-center molecules exposed to external electric DC or AC fields represents a computational challenge on several fronts.
On the one hand, there is the structure calculation of the molecule itself, which is now often tackled by quantum chemistry packages, usually employing
a basis that expands the orbitals in terms of Gaussians centered on the nuclei which accounts for the complete geometry. Recently, work has progressed
to allow the computation of Stark resonance parameters for small molecules within such an approach, both at the level of Hartree-Fock (HF) theory, as well as a correlated
coupled-cluster approach method involving singles doubles and perturbative triples (CCSD(T)) \cite{Jagau2014,Jagau2018}.
Photoionization of water molecules was recently
treated by a time-dependent R-matrix method~\cite{PhysRevA.102.052826}.

To deal with the resonance aspect one has to use an analytic continuation method, such as a complex absorbing potential (CAP), or complex exterior scaling.
The use of a CAP was demonstrated to yield accurate results for simple diatomic molecules, such as the hydrogen molecular ion, particularly when using
a correction scheme to remove the artefact of the CAP \cite{Riss_1993,Tsogbayar_2013}. Interesting observations can be made concerning the resonance properties
of the gerade vs ungerade lowest states of the molecule. The problem of putting the molecule into a low-frequency AC field can also be handled within the Floquet
approach in the Born-Oppenheimer approximation \cite{Tsogbayar_2013b}, as can the problem of harmonic generation by such fields \cite{Tsogbayar_2014}.

The CAP method requires the repeated calculation of complex eigenvalues along a trajectory determined by the strength parameter of the CAP (usually called $\eta$), which can be computationally demanding.
This can be avoided by an approach where exterior complex scaling (ECS) is introduced at a boundary, but a well-defined discontinuity in the wavefunction
and in its derivative is required \cite{se93}. The analogous strength parameter to $\eta$, is the scaling angle, which we will call $\xi$. Some pilot calculations of implementing the finite-element method (FEM) as proposed in Ref.~\cite{se93} were reported recently in a thesis \cite{pirkola2021exterior}. The results have been shown to be practically independent with respect to $\xi$ for a 1D hydrogen model, and to a certain degree also for the 3D hydrogen atom \cite{pirkola2021exterior}. We find the dependence of the resonance parameters (position and width) on $\xi$ much reduced for the FEM method with ECS as compared to the dependence of the CAP method on $\eta$ using the Wolfram Mathematica FEM solver, and an example is shown in Section~\ref{sec:acm}. 

The method can also be used as a perfect absorber for time dependent Schr\"odinger equation (TDSE) calculations \cite{PhysRevA.81.053845}. Furthermore, it was demonstrated that a time-dependent solution for the hydrogen model in a DC field using the ECS method can be well modelled by exponential decay, confirming the fundamental assumption of the time-independent ECS method \cite{pirkola2021exterior}. 

Our motivation for implementing a partial-wave approach is based in part on experience with TDSE calculations for proton-water molecule collisions in an atomic 
basis using a two-center basis generator method \cite{PhysRevA.80.060702,PhysRevA.85.052704},
which has led also to modelling of fragmentation of the molecule after the collision \cite{PhysRevA.85.052713}. 
These calculations were extended to projectiles with higher charges, and were
compared with experiment for $Li^{3+}$ projectiles \cite{PhysRevA.93.052705}.

Previous calculations were carried out on the basis of the single-center self-consistent field method of Moccia \cite{Moccia64}.
Stark resonance parameters were calculated for the valence orbitals  $1b_2$, and $1b_1$ \cite{PhysRevA.94.053413} by generating an
effective local potential from the simplified variational HF wavefunctions provided in Ref.~\cite{Moccia64}. For the $3a_1$ orbital the
dependence on the direction of the electric field was explored~\cite{Laso_2017}. These limited-geometry calculations involved solving a partial-differential equation eigenvalue
problem directly with a smooth exterior scaling approach, which is closer
in spirit to a CAP than to the presently employed ECS method with discontinuity
at the scaling radius. 

The effective potential was held constant for all field values in these resonance parameter calculations, and the field orientation
was in the molecular plane. The directional dependence focused on the field away from the oxygen atom along the center line passing in between the hydrogens,
and the reverse direction. 
We will focus on these two geometries in the present work, as well, since comparison for the orbital results can also be made with the work of Ref.~\cite{Jagau2018} for the entire molecule.

The present work is performed for a model potential that was developed to simulate the structure 
of the water molecule as described by the HF model \cite{PhysRevA.83.052704,ERREA201517}; the eigenvalues of the five molecular orbitals (MO) are also
given in Ref.~\cite{PhysRevA.102.012808}. We present an initial calculation of the resonance parameters of the water molecule in a DC electric field at the level of $\ell_{\max}=3$ with respect to the angular momentum basis functions. This corresponds to the level of Moccia's HF calculations \cite{Moccia64}.

The paper is organized as follows. In Sect.~\ref{sec:model} we introduce the methodology for the current work. In Sect.~\ref{sec:free} we present results for the field-free water molecule, in Sect.~\ref{sec:acm} we show an example
of a CAP trajectory along with ECS results, in Sect.~\ref{sec:density} we present a few figures of orbital densities for different applied fields, and in Sect.~\ref{sec:DCmo} the Stark resonance parameters for the three valence orbitals, $1b_1$, $3a_1$, and $1b_2$ are given in graphical form. 
In Sect.~\ref{sec:totalE} we introduce the problem of net (total) ionization of the molecule to compare with the HF and CCSD(T) results of Ref.~\cite{Jagau2018}. 
We make concluding remarks in Sect.~\ref{sec:conclusions}. The appendix in Sect.~\ref{sec:app} includes some details of the FEM implementation, along with tables of data for the figures shown in Sect.~\ref{sec:res}.
Throughout the paper, atomic units, characterized by $\hbar=m_e=e=4\pi\epsilon_0=1$, are used unless otherwise stated.

\section{Model}
\label{sec:model}
The model potential for the water molecule employed in the current work was developed 
to compute collisions of ions with water vapor, and was used
both in classical-trajectory calculations \cite{PhysRevA.83.052704,PhysRevA.99.062701,PhysRevA.102.012808} as well as a finite-difference approach to solving the TDSE \cite{ERREA201517}.
Of particular interest in those studies was the orientation dependence of capture and ionization cross-sections.
The model consists of a superposition of three spherical parts, representing the oxygen atom, and the two hydrogen atoms respectively using the
geometric arrangement predicted by HF calculations. 

The three parts aim to describe the static screening in the sense of an effective Hartree potential with
self-energy correction, i.e., the potential experienced by one of the electrons in the water molecule falls off as $-1/r$, as required.  Such a local potential method
could be compared to the density functional theory method of optimized effective potential 
(referred to as OEP or OPM). 
Calculations for such a model yield similar
MO eigenvalues for the valence orbitals (cf. Ref.~\cite{Goerling2007}). 
These models allow one to construct single-electron excited states, which is important for applications in collision physics, as well as laser excitation. 
More recently the classical-trajectory mean-field calculations based on the model potential were also tested for net electron capture by highly charged ions
from water vapor, and compared to independent-atom model calculations as well as experiment \cite{atoms8030059}.

The model potential can be written as
\begin{equation}\label{eq:pot0}
V_{\rm{eff}}=V_{\rm{O}}(r)+V_{\rm{H}}(r_1)+V_{\rm{H}}(r_2)\ , 
\end{equation}
\begin{align}\label{eq:pot}
\begin{split}
V_{\rm{O}}(r)=&-\frac{8-N_{\rm{O}}}{r}-\frac{N_{\rm{O}}}{r}(1+\alpha_{\rm{O}}r)\exp(-2\alpha_{\rm{O}}r)\ , \\
V_{\rm{H}}(r_j)=&-\frac{1-N_{\rm{H}}}{r_j}-\frac{N_{\rm{H}}}{r_j}(1+\alpha_{\rm{H}} r_j)\exp(-2\alpha_{\rm{H}}r_j)\ , 
\end{split}
\end{align}
where $\alpha_{\rm{O}}=1.602$, $\alpha_{\rm{H}}=0.6170$ are screening parameters, $r_j$ represents the electron distance from either proton ($j=1,2$), and 
the electron density `charge parameters' were chosen as $N_{\rm{O}}=7.185$ and $N_{\rm{H}}=(9-N_O)/2=0.9075$. The planar geometry of the water molecule
is given by the O-H bond length of $R_j=1.8$ a.u., and an opening angle of $105$ degrees. The three model parameters were adjusted to reproduce HF orbital energies \cite{ERREA201517}.

We make the following wavefunction ansatz in spherical polar coordinates:
\begin{equation} 
\tilde{\Psi}(r, \theta, \phi) = \sum_{l=0}^{\ell_{\max}} \sum_{m=-l}^{l}{\sum_{i,n}^{I,N}\frac{1}{r}c_{inlm}f_{in} (r) Y_l^m(\theta,\phi) } \ , 
\label{shexp}
\end{equation}
where the $Y_l^m$ are the usual (complex-valued) spherical harmonics, i.e., eigenfunctions of the orbital angular momentum operators ${\hat L}^2$ and ${\hat L_z}$. The functions $f_{in}$ are local basis functions on interval $i$ of the radial box and are polynomials up to $nth$ order \cite{se93}. 

%The computational problem can also involve solving coupled ordinary differential equations for the usual global functions $u_{lm}(r)=rR(r)$ which the $f_{in}(r)$ replace.
%Such an approach was used previously in the context of Dirac resonances in supercritical fields \cite{PhysRevA.84.032517} in order to capture features of a two-center
%collision between two bare heavy ions. 
Although we will use the FEM-ECS approach for the majority of the paper, an alternative approach was used to confirm the FEM-ECS results. It is based on propagating coupled differential equations for the radial functions as initial-value problems. It suffices to say that as a check, it agrees with the resonance parameters of the FEM-ECS method \cite{PhysRevA.81.053845,se93}. The numerical
comparison with this alternative method is discussed briefly in Sect. \ref{sec:acm}.

The oxygen part of the effective potential is straightforward to implement, as is the potential due to the external electric field, the centrifugal potential, and the kinetic energy term. The hydrogenic parts are handled in the following manner:
first we decompose the potential of one hydrogen atom displaced along $\hat z$ with distance $R_j$ using an expansion in terms of Legendre polynomials by making use of azimuthal symmetry for a single O-H problem:
\begin{equation} 
V_{\rm{H}}(r_j) \approx \sum_{\lambda=0}^{\lambda_{\max}}{V_\lambda(r) P_\lambda(\cos{\theta})}.
\end{equation}
The channel potentials $V_\lambda(r)$ are obtained by projecting $V_{\rm{H}}(r_j)$ as given in Eq.~(\ref{eq:pot}) onto Legendre polynomials.
It follows from selection rules that $\lambda_{\max}=2 \, \ell_{\max}$ is required. This implies that the truncated partial-wave expansion samples the effective potential
only up to some degree of accuracy, and thus a careful convergence study ought to be carried out, but this is not the subject of the present paper. Test calculations
have shown that the $\ell_{\max}=3$ calculations capture the salient features of the
Stark resonance problem for the water molecule.

The approximated hydrogen potential is rotated into place by means of the addition theorem of the spherical harmonics
\begin{equation} 
P_\lambda(\cos{\theta})=\frac{4\pi}{2\lambda+1}\sum_{\mu=-\lambda}^\lambda{Y_\lambda^\mu({\hat r'}){\bar Y}_\lambda^\mu({\hat r''})} \ ,
\end{equation}
where $\cos{\theta}={\hat r'} \cdot {\hat r''}$.  One then obtains two such expansions (one for each hydrogen atom), with the orientations
defined by $\theta'$ corresponding to one half of the opening angle (i.e., 52.5 degrees), and $\phi_1' = \pi/2 $ and  $\phi_2' = 3\pi/2 $ radians in accord
with the set-up described in Ref.~\cite{Moccia64} to place the molecule into the $y-z$ plane.

The truncated potential for one of the hydrogen atoms then becomes
\begin{equation} \label{eq:vh}
V_{\rm H}(r,\theta,\phi) = \sum_{\lambda=0}^{\lambda_{\max}} { \frac{4\pi}{2\lambda+1} V_\lambda(r) 
\sum_{\mu=-\lambda}^\lambda {Y_\lambda^\mu(\theta,\phi) 
{\bar Y}_\lambda^\mu(\theta',\phi_j)} } \ ,
\end{equation}
where the bar denotes complex conjugation. To form the matrix elements, first one sandwiches this potential with Dirac bra-kets between spherical harmonics, namely the
harmonic from the expansion of Eq.~(\ref{shexp}) of order $(l', m')$ and the harmonic onto which the resulting expression is projected, i.e, the harmonic of order $(l, m)$.

The $V_\lambda(r)$ are integrated in the radial parts of the matrix elements before the double sum in Eq. (\ref{eq:vh}) is applied, and the angular part of the matrix elements can be evaluated by means of Gaunt integrals (which make the selection rules clear, and result in $\lambda_{\max}=2 \, \ell_{\max}$
when applying $l'=l=\ell_{\max}$).

The inclusion of the external electric potential is very straightforward
and leads to couplings: for the field or force along $\hat z$ it will mix with $\Delta l = l-l' = \pm 1$ and $\Delta m = m-m' =0$, and therefore not mix the
$a$-states with the $b$-states of $\rm H_2O$. More clearly, the molecular orbitals $1a_1$, $2a_1$, and $3a_1$ do not mix with the other symmetries, i.e., $1b_1$ and $1b_2$.

To compute complex-valued resonance energies we apply exterior scaling with a discontinuity at the scaling radius $r_{\rm s}$ \cite{PhysRevA.81.053845,se93}.
One applies the transformation
\begin{equation}
r \to r_{\rm s}+(r- r_{\rm s})e^{i \xi} \ ,
\end{equation} 
where the scaling angle $\xi < \pi/2$ can be chosen large (close to its maximum allowed value), but as demonstrated in Sect.~\ref{sec:acm} the results are quite independent of it.
The choice of scaling radius is analogous to that of $r_{\rm c}$, but perhaps more importantly than for the CAP method it really should be in the range where
the effective potential (including the centrifugal potential) is a simple inverse-distance potential. This is to simplify the form of the final scaled potential. The kinetic energy operator acquires a factor of
$e^{-2 i \xi}$, and the wavefunctions under scaling should satisfy
\begin{equation}
\Psi(r_{\rm s}^-)=\Psi(r_{\rm s}^+) e^{i \xi /2} \qquad {\rm and} \qquad \Psi'(r_{\rm s}^-)=\Psi'(r_{\rm s}^+) e^{i 3\xi /2} \ , 
\end{equation} 
where $r_{\rm s}^-$ and $r_{\rm s}^+$ refer to the left- and right-hand sided limits of the scaling radius. This condition is easily implemented when solving a matrix problem using local radial basis functions, as the discontinuity can be put on a border between two sets of local radial basis functions. This is done by including an extra factor of $e^{i\xi}$ in the matrix elements outside of the scaling radius. This can be thought of as either the Jacobian for the scaling ($d\tilde{r}/dr$) or the combination of each wavefunction discontinuity applied to the basis functions. However, since the basis functions need not be scaled in principle, it is more satisfying to argue that this is just the Jacobian. This allows us to integrate over the unscaled variable with the measure $dr$, which is useful since the local basis functions are defined over it.

Calling the scaled radial variable $\tilde{r}$, it is not hard to see why the kinetic energy operator acquires negative arguments in the exponential, since,
\begin{equation}
\frac{\partial }{\partial \tilde{r}}\Psi(\tilde{r}) = \frac{\partial r}{\partial \tilde{r}}\frac{\partial \tilde{r}}{\partial r}\frac{\partial }{\partial \tilde{r}}\Psi(\tilde{r}) .
\end{equation}
Using the chain rule backwards the negative argument comes out in the next step,
\begin{equation}
\frac{\partial }{\partial \tilde{r}}\Psi(\tilde{r}) = e^{-i\xi}\frac{\partial }{\partial r}\Psi(\tilde{r}) .
\end{equation}
Again we are allowed to compute with respect to the unscaled variable. Since we use a symmetric form of the kinetic energy operator, the factor of two in the argument of the exponential comes out by the multiplication of the two single derivatives. Neither exponential is conjugated since only the left wavefunction in the matrix element was conjugated to begin with, and not the operator. Using integration by parts one can see how this works. 

We have presented the basics of the mathematical scheme we implement for computing the matrix problem for the Schr\"{o}dinger equation. The equation we solve for in differential form is,
\begin{equation}\label{eq:schro1}
\bigg[\frac{-1}{2}\nabla^2 - \sum_{i=1}^{3}\frac{Z_i(|\vec{r}_i|)}{|\vec{r}_i| } - F_z r  \cos(\theta)\bigg]\Psi = E\Psi \ ,
\end{equation}
where it is implied that the kinetic energy operator, and the $r$ variable in the potentials is scaled as discussed previously. The energy $E$ is complex-valued, with $E = E_{res}-i\Gamma /2$ defining resonance position and the full width at half-maximum $\Gamma$, which corresponds to the decay rate. 
Due to the geometry we use, a positive $F_z$ corresponds to a force which points towards the hydrogens, and a negative $F_z$ corresponds to a force which points towards the oxygen atom which is located at $r=0$.  

For our model each orbital responds individually to
the external field, i.e., with its own energy shift and decay rate.
There is no self-consistency since the potential is treated as a frozen
external potential. 
In the time-dependent approach one can attempt to take a mean field interaction
into account, and the responding orbitals can induce mutual interactions \cite{tel2013}.
Whether such mean-field interactions are working well in this context can
be a matter of debate. In collision problems where the time scales
are short time-dependent mean field theory appears to work quite well,
cf.~Ref.~\cite{PhysRevA.102.012808}, but for a DC problem such as the present
the situation is not so clear. In photoionization it leads to non-exponential
decay (the decay rate varies as a function of the ionization state of the
target described by mean field theory).
Therefore, we refer an investigation into this problem to future work.
The multi-electron system described by the Hartree-Fock approach (and more so in a correlated calculation, such as the
coupled-cluster method)
is dealt with as the decay of the entire molecule and not that of individual orbitals, i.e.,
it is based on the total energy~\cite{Jagau2018}. 
Comparison of the present model results at this level is discussed in
Sect.~\ref{sec:totalE}. 
%A relevant calculation using time-dependent DFT with exterior complex \cite{tel2013} scaling is a matter of interest in the conclusion as well. This is since they solve for the orbitals of the Kohn-Sham equation with a scaled Hamiltonian. However, they only describe the statistics of a general ionization event, i.e., they find a probability to measure the atom to not be ionized, along with a decay rate. 
\section{Results}\label{sec:res}

\subsection{Field-free molecular orbital energies}
\label{sec:free}

First we consider the eigenvalues of field-free water (i.e, water without an external field). In Table I the results listed by Jorge et al. are for the model potential from Eqs. (\ref{eq:pot0},\ref{eq:pot}) and calculated using Gaussian type orbitals (GTO) \protect \cite{PhysRevA.83.052704,PhysRevA.99.062701}. The results of Errea et al. solve this potential on a lattice with spacing $25/a_o$ (G25) \protect \cite{ERREA201517}. Also listed is the one-center self-consistent field (OC/SCF) results of Moccia \protect \cite{Moccia64} and their approximation which uses Slater type orbitals (STO) from Refs. \protect \cite{Laso_2017,PhysRevA.94.053413} labeled as SCF/STO. Also relevant is the optimized potential method (OPM), also referred to as
the optimized effective potential, which approximates the Hartree-Fock theory by a local potential \cite{Goerling2007}. The most accurate results of a multi-center STO approach is listed as WF IV (by their own naming) along with an experimental comparison from Aung, Pitzer, $\&$ Chan \protect \cite{pitzer}. More recent experimental results are listed as well \cite{LibreNix}.

\vspace{-0.15cm}
 \setlength{\tabcolsep}{9.7pt}
\bgroup
\def\arraystretch{0.7}
\begin{table}[!h]\label{table:first}
\begin{center}
 \begin{tabular}{ c c c c c c } 
 \hline
\hline
Molecular Orbital (MO) &$1a_1$&$2a_1$&$1b_2$&$3a_1$&$1b_1$\\

\hline
GTO \cite{PhysRevA.83.052704,PhysRevA.99.062701} & $ -20.25 $& $-1.194$&  $-0.737$& $-0.578$ & $-0.519$ \\ 
\hline
 G25 \cite{ERREA201517} &  NA & NA&  $-0.737$& $-0.568$ & $-0.518$ \\

\hline
 %GTO \cite{er15} &  NA& NA&  $-0.737$&$-0.578$ &$-0.519$ \\ 

%\hline
OC/SCF \cite{Moccia64}&$-20.5249$ & $-1.3261$& $-0.6814$ &$-0.5561$  &$-0.4954$\\
\hline
 SCF/STO \cite{Laso_2017,PhysRevA.94.053413} &  NA& NA&  $-0.682$& $-0.557$
 &$-0.497$ \\ 

\hline
 
  OPM \cite{Goerling2007} & $-18.9722$ & $-1.1783$& $-0.7127$& $-0.5829$& $-0.5091$\\
\hline
 WF IV \cite{pitzer} & $-20.5654$& $-1.3392$& $-0.7283$& $-0.5950$ & $-0.5211$\\ 
\hline
 Experiment \cite{pitzer} &  NA& NA&  $-0.595(11)$&$-0.533(11)$ &$-0.463(4)$ \\ 
 \hline
 Experiment \cite{LibreNix} & NA & NA &$-0.68$ & $-0.54$ & $-0.46$\\
\hline
 $\ell_{\max}=3$ (present) & $-21.4228$&$-1.1849$&$-0.7099$  &$-0.5718$&$-0.5213$\\ 
\hline
\hline
\end{tabular}

\caption{Energy eigenvalues of free water in atomic units. Our work is listed at the bottom. References are listed as well for other works. Note that OPM \cite{Goerling2007} yields less bound $1a_1$ and $2a_1$ energies, since they do not correspond to HF eigenvalues.}
\end{center}

\end{table}
 
The comparison with experiment shows some deviations, with theory giving deeper binding for the outermost orbitals. The experimental values are for single ionization, and the difference is due to orbital relaxation and electron correlation \cite{LibreNix}. The electron correlation for the neutral molecule and the ionized states are obviously not the same.
%The comparison with the values obtained by the authors who designed the potential using a Gaussian orbital based quantum chemistry program reveals
%that the outermost orbital $1b_1$ with its density perpendicular to the molecular plane is fast, because the density has limited overlap with the hydrogen
%atoms. 
Our two outermost energies fall slightly below the quoted value in Ref.~\cite{ERREA201517}, while the deepest valence orbital ($1b_2$) falls slightly above, the latter on account
of a lack of convergence in the partial wave expansion. 
Given that we are solving for a local model potential
comparison should be made with a local potential method, namely the OPM method (Ref.~\cite{Goerling2007}), and it is very good. 
The $1b_1$ orbital is referred to as the highest occupied molecular orbital, and Koopmans' theorem in HF theory \cite{Plakhutin2018} can be carried over to DFT methods
which have a correct asymptotic form of the effective potential. This has been investigated for molecules \cite{Tsuneda2010}. 

Since there are multiple valence orbitals in water one may ask for an `extension' of Koopmans' theorem by assuming the ionization energies are still close to the negative orbital energies.
What we mean by that is that different ionic states can be created for 
the water molecule: removal of one electron from a particular valence shell
leads to a different singly ionized water molecule. The question arises whether
on can connect the energy difference between the ground states of the ionized $\rm H_2O^{+}$ and the neutral $\rm H_2O$ to a particular valence orbital energy accurately. The situation is quite different from what is allowed for an atom.

The experimental ionization energies of water can be deduced from photoelectron spectroscopy, but they are complicated by
vibrational level structures and are significantly broadened; quoted values for the three valence orbitals correspond to ionization energies of $0.68, 0.54, 0.46$ respectively \cite{LibreNix}. An accurate determination of the $1b_1$ and $3a_1$ ionization spectra involves resolving rotational levels, as well \cite{Page1988}. 
More recent work is in progress on the basis of (e, 2e)
collision spectroscopy \cite{NixonWeb}.
%\footnotetext{These values are digitally extracted from %plots from the works of Arias Laso %\cite{Laso_2017,PhysRevA.94.053413}.}

\subsection{Analytic continuation methods}\label{sec:acm}

In this brief section we contrast the present FEM-ECS results against
another analytic continuation method, namely the complex absorbing potential (CAP)
approach which was used, e.g., previously in computations for the
Stark resonance problem in the molecular hydrogen ion~\cite{Tsogbayar_2013}. In a CAP method one artificially
modifies the Hamiltonian by adding a complex absorber.
The results depend on $\eta$, which is the strength parameter of the 
CAP, and a complex trajectory emerges.

In Fig. ~\ref{fig:cap} we show such a trajectory for 
the $3a_1$ eigenvalue using blue $\textcolor{NavyBlue} \bullet$ for the case of $F_z = 0.1$ a.u. and a range of $\eta$ values, $\eta \in [0.008, 0.06]$, spaced by $\Delta \eta = 10^{-3}$. The calculations were performed using a numerical differential
equation FEM eigensolver for homogeneous problems ({\bf NDEigenvalues} in Wolfram Mathematica). The trajectory is smooth for larger values of $\eta$, but deteriorates rapidly at small $\eta$ (on the left in Fig.~\ref{fig:cap}).

One can find a so-called stabilized value on the trajectory, and also
apply the perturbative correction method by Riss and Meyer \cite{Riss_1993} to remove the artefact of the CAP.
Applying this correction at first and second order results in
two closely spaced points (yellow $\textcolor{Dandelion} \bullet$ and purple $\textcolor{Purple} \bullet$) visually `below' the trajectory. Similar results
could also be obtained by Pad\'e extrapolation to $\eta=0$ of the trajectory from the well-behaved part of the trajectory, i.e., the `good' side
relative to the stabilization point.

\begin{figure*}[!h]
\centering
\hspace*{-1.5cm}
\includegraphics[width=550pt]{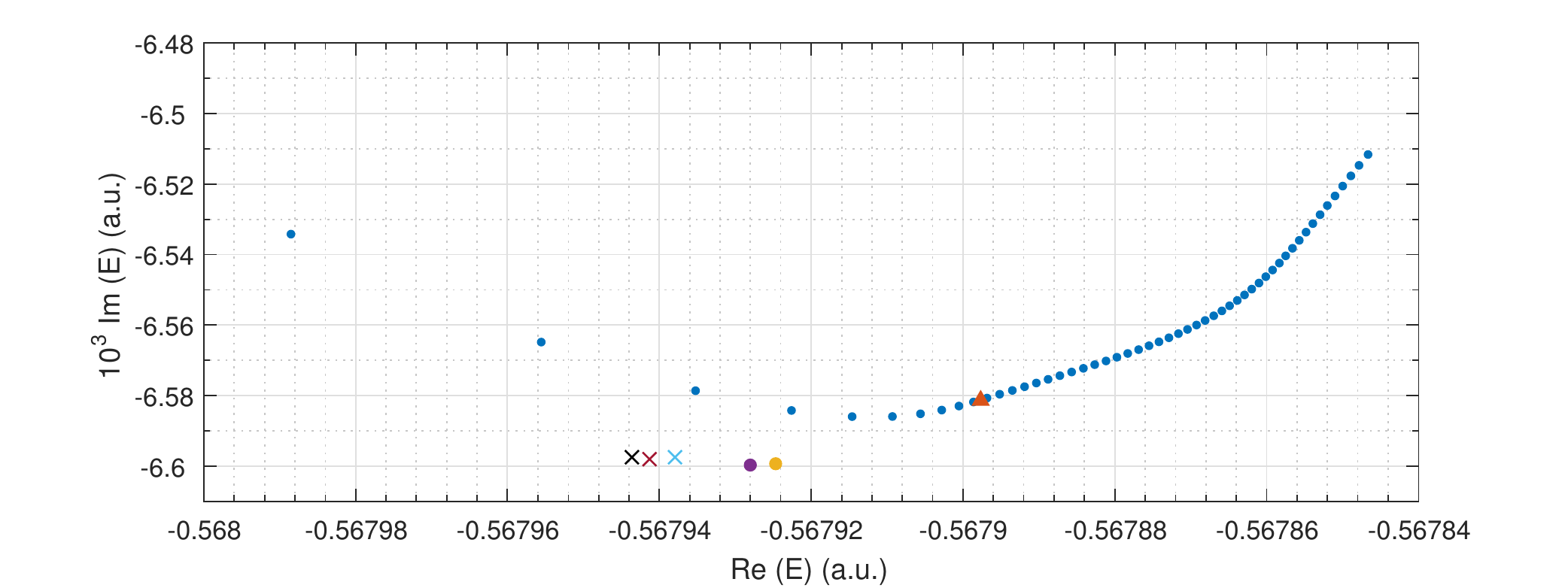}
\caption{\label{fig:cap}
The complex trajectory of $3a_1$ MO eigenvalues at $F_z = 0.1$ a.u. for the CAP method, along with ECS results. The major grid lines on both axes show variation at the $10^{-5}$ level, which illuminates the variation of complex eigenvalues for different solution parameters. The CAP trajectory is marked by blue $\textcolor{NavyBlue} \bullet$, for $\eta \in [0.008,0.06]$. The CAP stabilization method result is shown by the red $\textcolor{BrickRed} \blacktriangle$. The yellow $\textcolor{Dandelion} \bullet$ and purple $\textcolor{Purple} \bullet$ are the result of the first and second order CAP corrective schemes. The cyan $\textcolor{cyan} \times$ symbol marks the ECS result with $\xi \approx 1.4$ rad using a coupled differential equations method. The red $\textcolor{BrickRed} \times$ marks the FEM-ECS with 180 integrator nodes and $\xi =0.5$ rad. The black $\textcolor{black} \times$ marks two points, for $\xi=0.5$ and $\xi \approx 1.4$ rad, both with 120 integrator nodes.}
\end{figure*}

Looking at the trajectory and the coalescence of the stabilized value
($n=0$ in the correction scheme) and the perturbatively corrected 
values at order $n=1,2$ one observes for all the CAP results shown in Fig. ~\ref{fig:cap} agreement at the level of three decimal places.

The ECS method, on the other hand, practically shows no dependence
on its scaling parameter $\xi$, but may show some dependence on the
accuracy of integration in the FEM-ECS approach. As a note, for the FEM-ECS method we have used 120 integrator nodes for all results given in the tables.
%outside of Fig. \ref{fig:cap}. 
We confirmed the 
accuracy of these results (four decimal places) by an alternative
implementation of ECS using a coupled-differential equations approach listed as the cyan $\textcolor{cyan} \times$. We note that the ECS results still have a small issue as far as convergence is concerned, but certainly for the width parameter $\Gamma$ an accurate result has been obtained. The main point to be made is that a direct
calculation within the ECS framework yields an answer, and no trajectory computation is required.

Furthermore, in particular consideration of the coupled-differential equation approach for ECS, we can make a further comparison with our FEM-ECS results. For example, for the field strength of $F_z = 0.08$ a.u., the resonance positions of $1b_2$ agree up to four significant digits, and the widths up to three significant digits. The resonance positions of $1b_1$ agree up to nine significant digits, and the widths agree up to five significant digits. In other words, the worst agreement still captured the same order of magnitude, along with the first three digits used in scientific notation. Care was taken to make sure that the calculations were similar, but differences in methodology and basis choice (and size) remain as explanations for the disagreements. 
\subsection{Density plots for the molecular orbitals}\label{sec:density}
%\begin{figure}[b]
%\centering
%\begin{subfigure}{.49\textwidth}
%    \centering
%    \includesvg[width=0.8\textwidth]{resm3a1L3.svg}
%\end{subfigure}%
%\begin{subfigure}{.49\textwidth}
%    \centering
%    \includesvg[width=0.8\textwidth]{resm1b2L3.svg}
%\end{subfigure}
%\begin{subfigure}{.49\textwidth}
%    \centering
%    \includesvg[width=0.8\textwidth]{res03a1L3.svg}
%\end{subfigure}%
%\begin{subfigure}{.49\textwidth}
%    \centering
%    \includesvg[width=0.8\textwidth]{res01b2L3.svg}
%\end{subfigure}
%\begin{subfigure}{0.49\textwidth}
%    \centering
%    \includesvg[width=0.8\textwidth]{resp3a1L3.svg}
%\end{subfigure}
%\begin{subfigure}{0.49\textwidth}
%    \centering
%    \includesvg[width=0.8\textwidth]{resp1b2L3.svg}
%\end{subfigure}
%\caption[short]{A beautiful, well written caption}
%\end{figure}

We present some graphical results to demonstrate
the quality of the partial-wave expanded orbitals which combine
an accurate representation for the radial part of the wavefunction with
a limited superposition of spherical harmonics to represent the angular portions of the wavefunction. We begin with the MO $3a_1$ which
is composed of even-$m$ harmonics and the bonding $1b_2$ orbital
composed of odd-$m$ harmonics: both in the field-free case,
and in the cases where the field is aligned (or counter-aligned)
with the chosen $z$ axis.

\clearpage

\begin{figure*}[!h]
\centering
\vspace*{0cm}
\includegraphics[width=325pt]{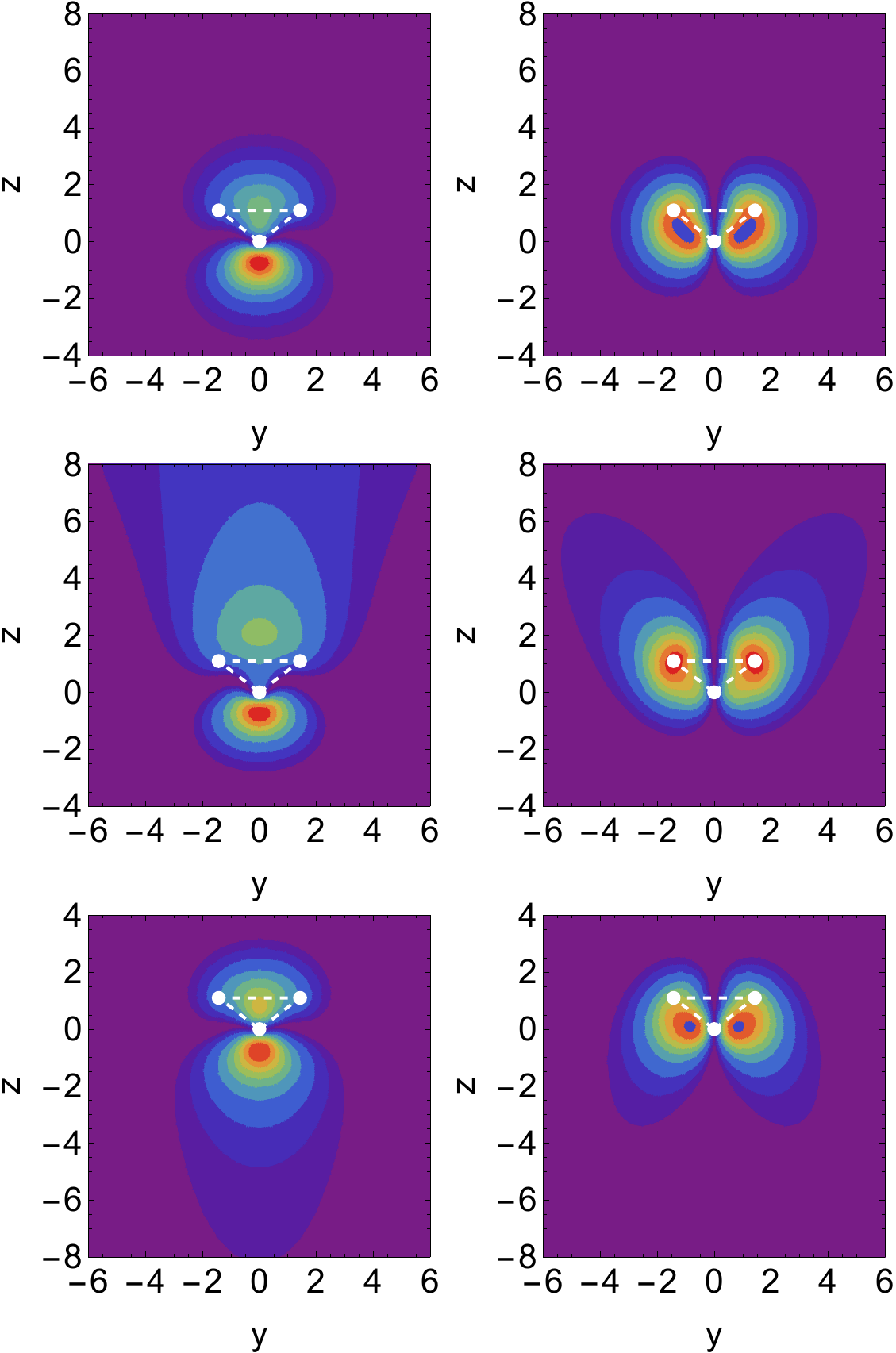}
\caption{\label{fig:grid2by3}  All axes are in atomic units. Contour plots of normalized probability densities for the $3a_1$ MO (left column) and the 
bonding $1b_2$ MO (right column), over the molecular plane (denoted $y-z$) with the
oxygen nucleus and the protons
marked by a white triangle. 
Top row: field-free case,
middle panel: for a field 
pushing electrons away from oxygen past the protons; bottom panel: opposite field direction. The field strength is $0.1$ a.u. for the case of $3a_1$,
and $0.2$ a.u. for $1b_2$.
The contour values (starting from the outside)
are $0.005, 0.01, 0.02$ followed by a regularly
spaced sequence of intervals of $0.02$ a.u..
}
\end{figure*}
\clearpage

Fig.~\ref{fig:grid2by3} above shows the probability densities in the molecular plane,
labeled as $y-z$. The protons are located at $y \approx \pm 1.428$
and $z \approx 1.096$ atomic units. Looking at the density distributions in the left column we observe how the probability density changes 
from the field-free case (top row) to a situation in which 
electrons flow away from oxygen past the protons mostly in the 
direction along the $z-$axis for one field direction (middle row),
and in the opposite direction for the reversed field (bottom row).
The calculation for the case with decay was obtained using the CAP
method. In the (left) middle panel one can see the relevant orbital for the stabilized eigenvalue value along the trajectory
shown in Fig.~\ref{fig:cap}.

For the field in the opposite direction (bottom left panel) the decay
rate is smaller by about a factor of four
(the magnitude of the imaginary parts is 0.00658 versus 0.00165 atomic units). 
One can again observe 
a probability density describing electron flux centered along the $z-$ axis, this time, toward the oxygen atom and beyond. At the same time, an increase in electron density occurs in-between
the two protons. One can see for both force orientations along $z$, that the density lobe on the opposite side of the axis (say, the lobe on the positive side of the $z-$axis when the force is in the negative $z$ direction) becomes smaller and has a higher density of contour lines.

For the bonding $1b_2$ orbital which is shown for a doubled field strength in the right column ($F_{0,z}=\pm 0.2$ a.u.) a different
behavior can be observed. This MO has probability density concentrated
between the oxygen and proton nuclei (top right panel). 
With the field pushing electrons out, away from oxygen, past the protons (middle right panel), one can observe directional flow somewhat along
the bond axis with deformed humps in the probability distribution.

For the field pushing in the opposite direction (bottom right panel)
emission is favored again in an outward direction.
The imaginary parts of the energy eigenvalues have magnitudes of
0.00306 and 0.00169 atomic units respectively for the two cases. This is consistent
with what is shown in the probability plots, i.e., a stronger
deformation towards larger distances in the middle versus the 
bottom panel.

\clearpage

\begin{figure*}[!h]
\centering
\vspace*{0cm}
\includegraphics[width=160pt]{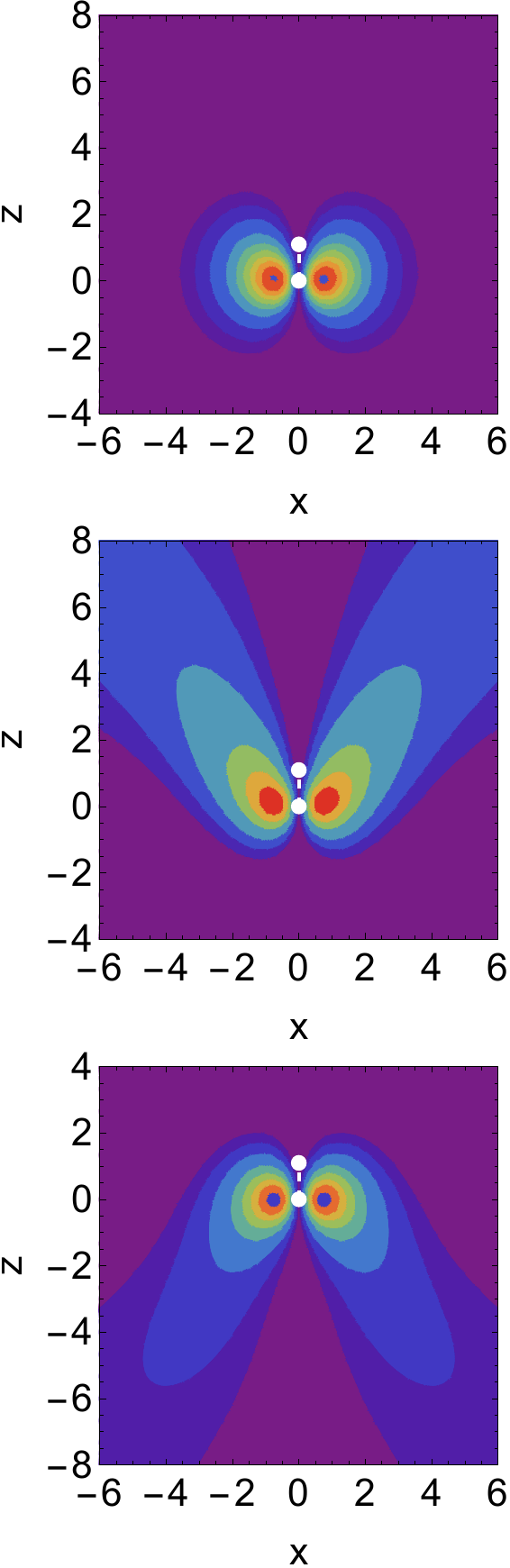}
\caption{\label{fig:grid1by3}  All axes are in atomic units. Contour plots of the normalized probability density for the (highest occupied) $1b_1$ MO, over the $x-z$ plane (perpendicular
to the molecular plane). 
White dots mark the oxygen nucleus and the proton location's projection.
Top panel: field-free case,
middle panel: for a field with strength $0.2$ a.u.
pushing electrons away from oxygen past the protons; bottom panel: opposite field direction. The contour values (starting from the outside)
are $0.005, 0.01, 0.02$ followed by a regularly
spaced sequence of intervals of $0.02$ a.u..
}
\end{figure*}
\clearpage

In Fig.~\ref{fig:grid1by3} above the probabilities for the `lone' MO $1b_1$ are shown.
It is the least bound MO (the so-called highest occupied MO), with probability
density concentrated perpendicular to the molecular plane, which
in our notation is the $y-z$ plane. Again, we observe an asymmetry
for emission when the external force field is aligned or counter-aligned with the $z-$axis. As was done for the $1b_2$ MO the field strength
was doubled compared to the $3a_1$ case to obtain stronger
decay rates.
The imaginary part magnitudes of the stabilized CAP eigenvalues
are approximately
0.0144 and 0.00816 atomic units respectively for the cases shown in the middle vs bottom panel.

To summarize this section we note that the truncation of the
partial-wave expansion at $\ell_{\max}=3$ does not prevent us from 
observing interesting features in the resonance states which
describe very distinctive patterns of ionized electron flux.
The density plots also display geometrical displacements of the bound parts which are associated with the external field, and these lead to interesting features in the Stark shifts as a function of field orientation and field strengths. Such features were observed recently in HF and CCSD(T) calculations~\cite{Jagau2018}, but were absent
in single-center model potential calculations~\cite{Laso_2017}.
%\clearpage
\subsection{Resonance positions and resonance half-widths for molecular orbitals of $\rm H_2 O$}
\label{sec:DCmo}

We now present results for Stark resonance parameters using the FEM-ECS method. In the first subsection the figures are labelled by the molecular orbital (MO) corresponding to the field-free molecule. Note that due to spherical harmonic mixing caused by the DC field, the orbitals obtained for the resonances do not truly correspond to the field-free orbitals. Nevertheless, the progression of the orbital energies is traced by extracting the data from the code which corresponds to a smooth transition of the energies from the field-free molecule to the given field strength. The data is plotted, for the resonance positions (the real parts, or $\mathfrak{Re}~\rm (E)$) and half-widths (the negative of the imaginary parts, or $-\mathfrak{Im}~\rm (E)$). The half-widths are labelled as $\Gamma/2$, where $\Gamma$ is the decay rate, i.e, the inverse of the decay time constant $\tau$. 
% this data has moved to the appendix, so we should not discuss it here
%In the second subsection, our data are compared with other sources. In these tables our data is truncated to make comparison with the other sources easy. In general, we carry digits up to and including the maximum number in any given table. The last digit is rounded accordingly.

To obtain the results of the current work we use a radial box from 0 to 24.3 a.u. with 24 intervals and 11 radial basis functions (polynomial type) per interval. The angular momentum basis is cut at $\ell_{\max} = 3$, to include $\ell=0,1,2,3$ and all associated $m$ values. Due to our particular choice of orientation of the water molecule, the $n a_1$ orbitals have even-$m$ basis functions, while only odd $m$ values contribute for the $1b_1$ and $1b_2$ orbitals. 

For the resonance parameters, the scaling radius is set at 16.2 a.u., and the scaling angle is set at 81 degrees or $\sim$ 1.4 rad, close to the maximum allowed value (90 degrees). The scaling radius is chosen sufficiently large such that the potential energy experienced by the electrons behaves like $-1/r$. The role of the scaling angle
$\xi$ is to ensure that at the box boundary (24.3 a.u.) the demand of a vanishing
wavefunction does not introduce a significant artefact.

\begin{figure*}[!h]
\centering
\hspace*{-0.5cm}
\includegraphics[width=500pt]{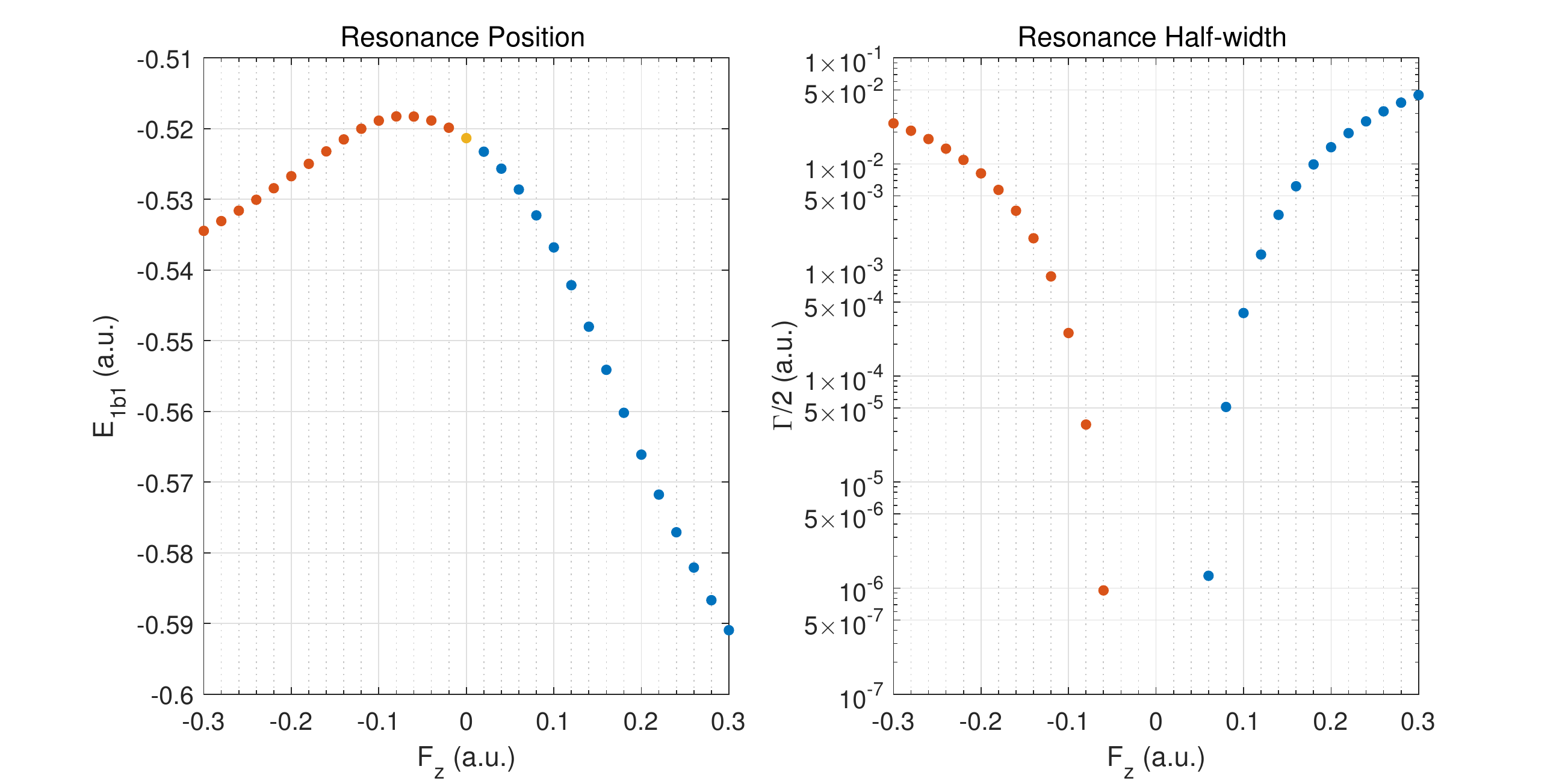}
\caption{\label{fig:1b1scrin}
Resonance positions (left panel) and half-widths (right panel) of the $1b_1$ valence orbital of $\rm H_2O$ in an external DC electric field. The sign of the values of $F_z$ (horizontal axis) points in the direction of the force experienced by an electron. The energy position of the free water molecule is given by the yellow marker. The orange markers are for points with the negative values of $F_z$ while the blue markers indicate results for $F_z > 0$. As explained in the text, positive $F_z$ values corresponds to the force driving the electrons from the oxygen atom towards the hydrogen atoms and beyond; negative values correspond to the force driving the electrons from the hydrogen atoms towards the oxygen atom and beyond.}
\end{figure*}
In Fig.~\ref{fig:1b1scrin} the results of complex-valued eigenenergies are shown for the outermost valence orbital $1b_1$: the left panel
shows the energy position of the resonance, and the right panel the negative of the imaginary part, which represents
the half-width at half-maximum of the Breit-Wigner resonance. When the electric force is pointing towards the hydrogen atoms, away from the oxygen ($F_z > 0$), the electrons are pushed outward and towards the protons, thereby lowering their eigenenergy considerably (binding being represented
by the magnitude of the real part). When the electric force experienced by the electron is pointing towards the oxygen atom ($F_z < 0$) the electrons are attracted towards it, causing initially a weak
decrease in binding. Eventually the attraction to oxygen increases their binding as the field magnitude increases. 

We notice a few remarkable features in these results: first of all, the DC Stark shift is quite asymmetrical about $F_z=0$. For negative values,
i.e., the external field pushing electron density away from the protons
first raises the orbital energy at weak field strength due to the attractive potential, but then the trend turns around as the field
strength increases and drives electron density towards the oxygen atom.

This is accompanied by a rise in exponential decay rate as shown
in the right panel (orange markers). At weak fields there is the
tunneling regime with a very strong dependence of the resonance width
on the field strength. As the field strength increases one observes
turnover, and approach towards an over-barrier ionization regime.

For the opposite field direction (blue markers), i.e., $F_z>0$,
a monotonic lowering of the orbital energy is observed. Nevertheless,
the decay rate shows a similar growth pattern with field strength,
and, in fact, exceeding the rate observed in the opposite direction
once the over-barrier regime is reached by about a factor of two.

This orbital has a probability distribution which is perpendicular
to the molecular plane formed by the three nuclei. One might therefore
assume that the DC shifts are not too strong, and for moderate fields this is indeed the case. Concerning the comparison with a previous
model calculation \cite{Laso_2017} which used an effective potential derived from
a single-centre HF solution and which was limited to $F_z>0$, we observe that the trends are similar. The DC shift is more pronounced
in the present calculation, but the behavior of the decay rate is quite comparable.

\begin{figure*}
\centering
\hspace*{-0.5cm}
\includegraphics[width=500pt]{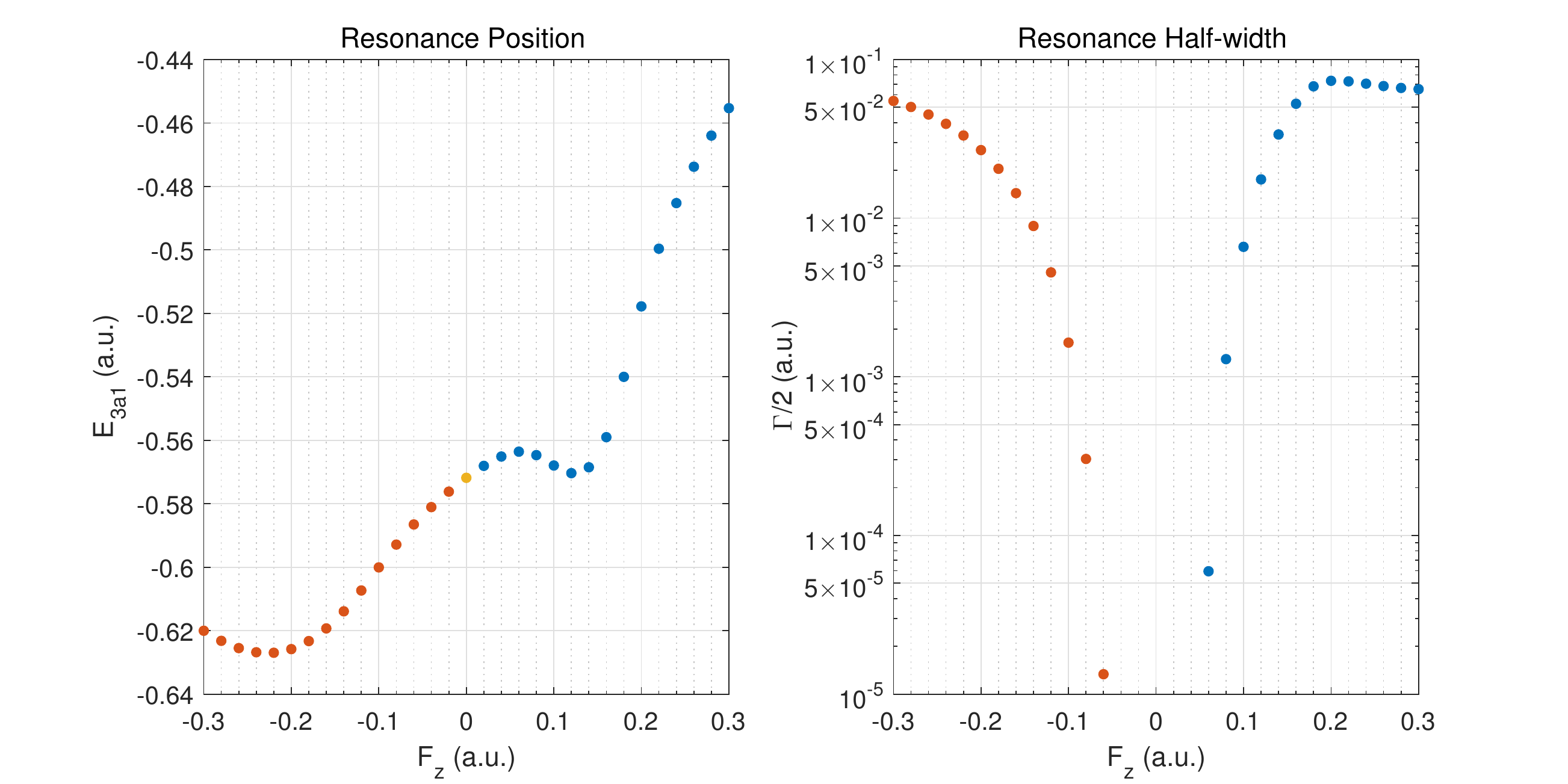}
\caption{\label{fig:3a1scrin}
Resonance positions (left panel) and half-widths (right panel) of the $3a_1$ valence orbital of water in an external DC electric field.}
\end{figure*}
Fig.~\ref{fig:3a1scrin} shows results for the valence orbital $3a_1$ and these are quite remarkable: the resonance position is not changing in
a monotonic fashion when the electric force pushes electrons out on the side where the hydrogens are located. The Stark shift becomes in fact very small
at some critical field strength. There are two undulating (or wave) features, one on each side of the $F_z$ axis. A similar feature was reported by Jagau in their molecular Stark shift results based on the total energy for the $\rm H_2O$ molecule \cite{Jagau2018}.
This is discussed in Sect.~\ref{sec:totalE}.

Comparing these results with the previous work based on the single-center
SCF potential with assumed azimuthal symmetry~\cite{Laso_2017} indicates that the present work
has more geometric flexibility in the resonance solution, and therefore
shows more prominent features in the behavior of DC shift and decay rate $\Gamma$ as a function of field strength and field orientation.
The work of Ref.~\cite{Laso_2017} showed monotonic behavior in the DC shift with a decrease in orbital energy.
At small $|F_z|$ the widths from these calculations
are similar to the current results, but with increasing field strength
the decay rate in Ref.~\cite{Laso_2017} becomes higher for the case where ionized electrons move
away in the direction from oxygen past the two protons by
about a factor of two at $|F_z|=0.3$ a.u..  
 Note that the definition
of the sign of $F_z$ is reversed in the present work with respect to Ref.~\cite{Laso_2017} such that it refers to the force experienced by the electron.

To summarize the comparison between the present results and those
of Ref.~\cite{Laso_2017} we may state that the inclusion of the 
dependence in the effective potential on the azimuthal angle
$\phi$ is important and plays a role in the rich 
behavior observed for the DC Stark shift of the $3a_1$ orbital.
%is apparent on both sides of the axis. At face value, their A configuration does not show a net wave. This means at least that the wave behaviour of $3a_1$, if it exists for their orbital, is hidden in their A configuration. In the conclusions we explore a comparison with Jagau. In the following tables we show comparisons for the eigenvalues of free water from a number of references and the resonance parameters in the figures above with Jagau. The resonance half-widths also show stabilization for large positive force values (or $F_z$).

\begin{figure*}[!h]
\centering
\hspace*{-0.5cm}
\includegraphics[width=500pt]{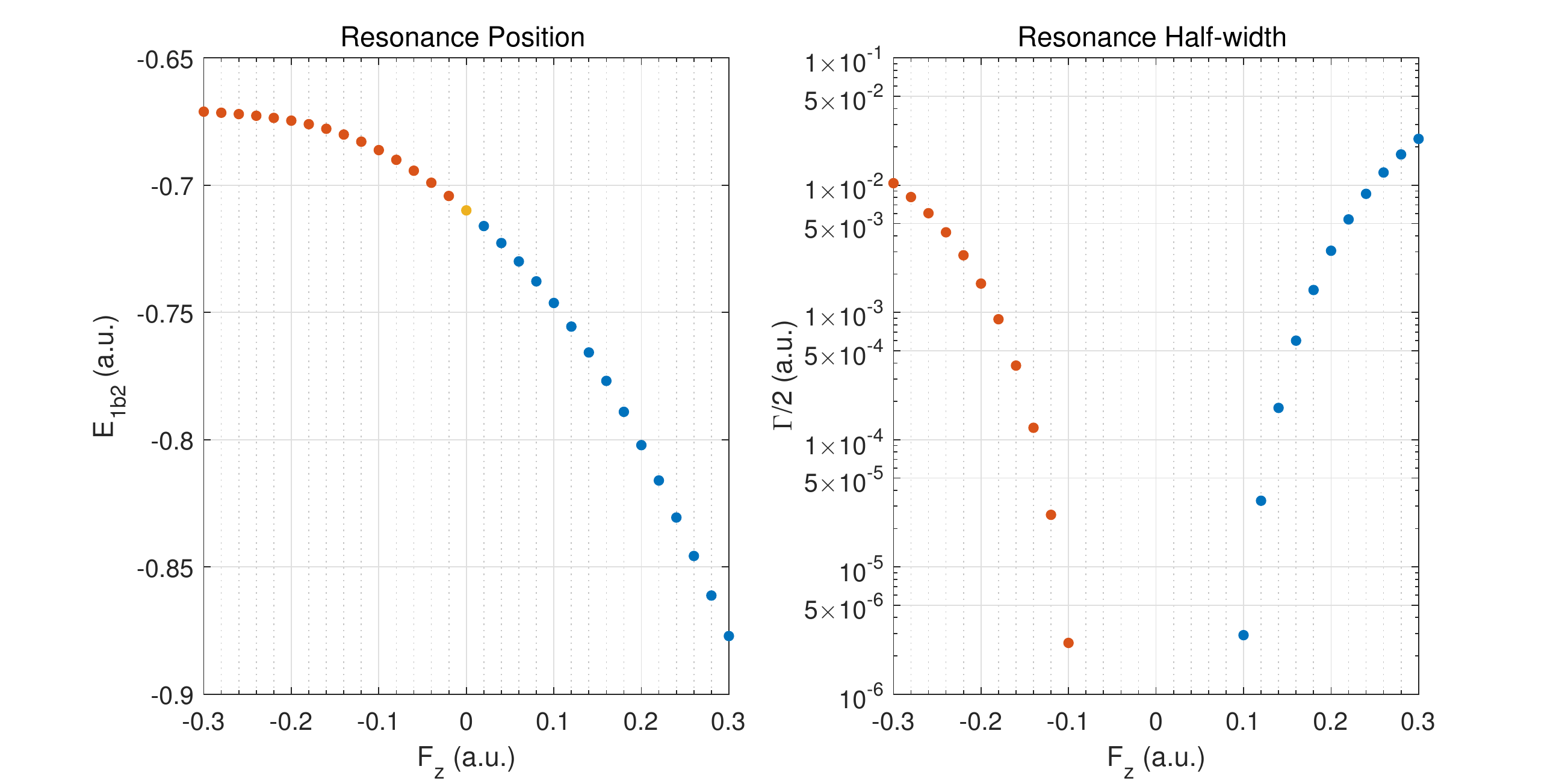}
\caption{\label{fig:1b2scrin}
Resonance positions (left panel) and half-widths (right panel) of the $1b_2$ valence orbital of water in an external DC electric field.}
\end{figure*}
In Fig.~\ref{fig:1b2scrin} above the results are shown for the bonding orbital $1b_2$, which has the strongest binding
of the three valence orbitals. We observe that when the electric force points towards the oxygen atom, the electrons actually become slightly less bound, and the opposite is true when the density is pushed
towards the protons. 
Despite the negative Stark shift the ionization rate is stronger
for $F_z>0$. We also observe that for this deeply bound valence orbital
the decay rates are smaller in comparison, and for the given field strength range do not get far beyond the tunneling regime.

We would like to remark on the fact that all three reported orbital calculations were performed independently in a frozen model potential,
i.e., no self-consistency based on the orbital densities was attempted.
A fully self-consistent calculation, in the sense of DFT, such as, e.g.,
the OPM approach will cause deviations for the energy shifts and decay
rates.

\subsection{Comparison with work based on total energy}
\label{sec:totalE}

We now connect our model potential work with
the work of Jagau, which is based on the total energy of the
molecule, and which was performed at the level of HF theory, as well
as a correlated method, namely coupled-cluster theory (CCSD(T))~\cite{Jagau2018}. It is an attempt only, because the
model potential does not lead to an obvious definition of the
total energy.

In HF theory one can express the total energy as
\begin{equation} 
E_{\rm HF} = \frac{1}{2}
\left(\sum_j\epsilon_j + 
\langle \psi_j| V_{\rm nuc}|\psi_j\rangle \right )  ,
\label{eq:Ehf}
\end{equation} 
where the sum is over occupied orbitals.
The sum of orbital energies may give only one half of the total energy
or less, and this is made up by the interaction energy matrix element
with the external potential. The reason for the required correction beyond the sum of eigenvalues is that the latter double-count the
direct and exchange contributions from the electron-electron repulsion.

In principle, we could calculate such an energy expression,
and derive a single decay rate. For now we limit ourselves
to a direct sum over orbital energies, which also includes a factor
of two to account for the spin degeneracy.
A decay rate calculation of this sort corresponds to net ionization,
i.e., ten electrons are participating in the decay of the molecule. 
Of course, this total decay rate is dominated by the contributions from
the valence orbitals discussed in Section~\ref{sec:DCmo}, but
for the DC Stark shift contributions are also coming from the inner
orbitals. The prescription for the total (or net) decay rate which 
is based on the sum of complex-valued orbital energies is consistent
with what has been used in the time-dependent domain by employing a straightforward multiplication of orbital probabilities and analyzing a total system survival probability~\cite{tel2013}. The method of Telnov et al. \cite{tel2013} results in a sum over orbital decay rates for the net decay rate when applied to our orbitals, due to the multiplication of exponential time-dependencies in the squared probability amplitudes. This is defined by,
\begin{equation}
    P_s(t) =\prod_{n,\sigma} \int_{0}^{r_s}d^3r|\Psi_{n,\sigma}(\vec{r},t)|^2,
\end{equation}
where the orbital wavefunctions are given by $\Psi_{n,\sigma}(\vec{r},t)$, along with a decay rate defined as,
\begin{equation}
\Gamma(t) = - \frac{d}{dt}{\rm ln} P_s(t).
\end{equation}}

\begin{figure*}[!h]
\centering
\hspace*{-0.5cm}
\includegraphics[width=500pt]{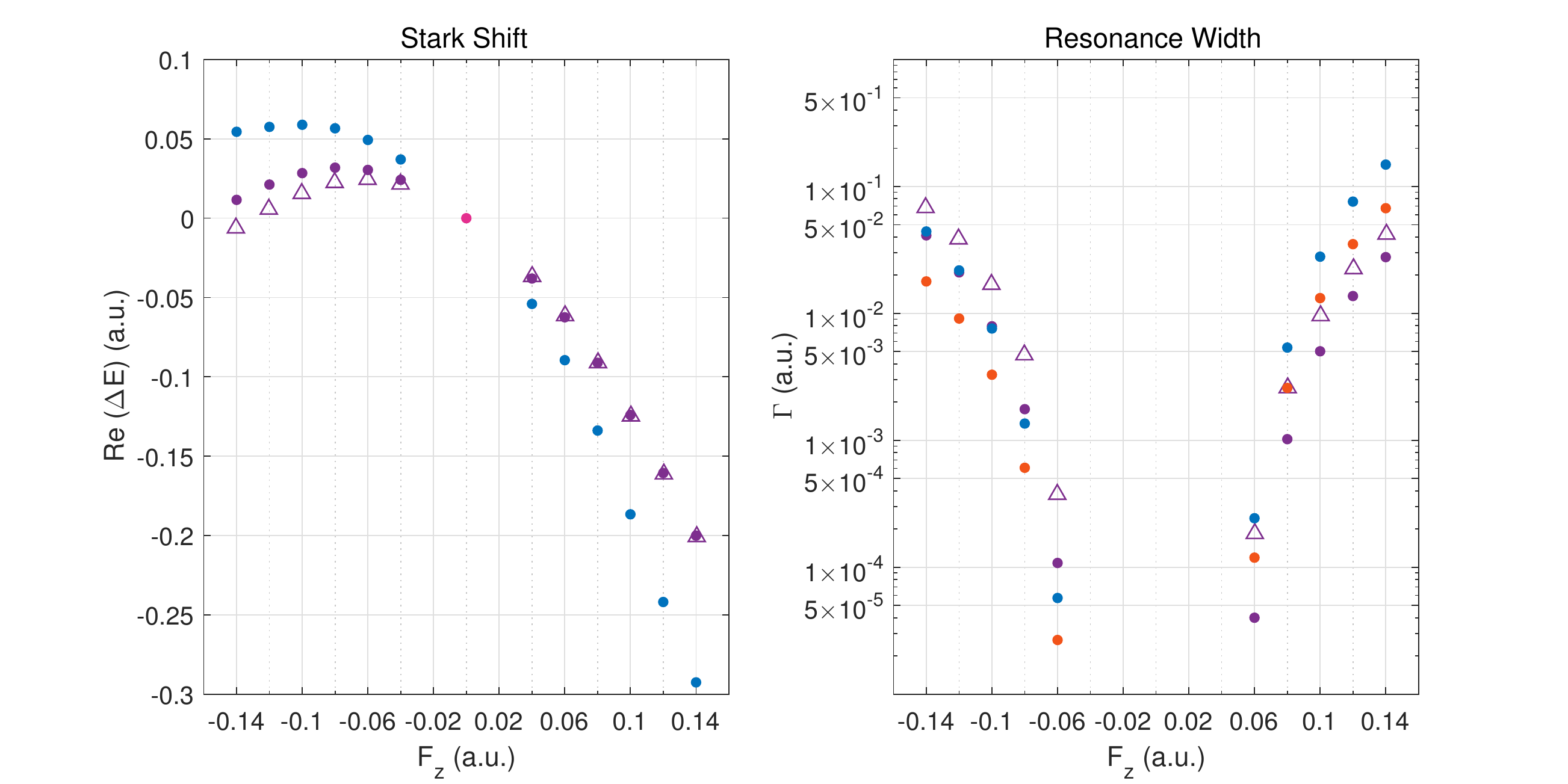}
\caption{\label{fig:jagau}
The net resonance positions and widths for Jagau \cite{Jagau2018} and our direct sum method. The blue dots indicate our direct sum method, and the purple dots and triangles indicate the HF and CCSD(T) methods of Jagau \cite{Jagau2018}. The red dot in the left panel is the reference by definition; $\Delta E =0$ for $F_z = 0$. The orange dots on the right panel indicate the widths of the orbital $3a_1$. }
\end{figure*}

Fig.~\ref{fig:jagau} shows in the left panel the Stark shift, and in the right panel the full width (decay rate in atomic units). It is compared with the HF and CCSD(T) results, and displays at least a similar
pattern. For the Stark shift it appears as if the $1b_2$ orbital provides the dominant contribution, as the result is similar in shape. 
This bonding orbital is 
relatively deeply bound and the response of the resonance position to the external field is
strongest among the valence orbitals as can be seen
from the scale in the left panels of Figs.~\ref{fig:1b1scrin}~$-$~\ref{fig:1b2scrin}.
The Stark shift from the present
work based on the sum of orbital
energies
acquires a strong contribution from the inner 
$2a_1$ orbital, 
as can be seen in Table~\ref{tab:totE} in the Appendix, i.e., in Sect.~\ref{sec:HFcomp}.

The HF results of Ref.~\cite{Jagau2018} show a strong turn-around in the Stark shift: as the field strength reaches $F_z=-0.15$ a.u. it
approaches zero (with CCSD(T)
giving a zero crossing at about
$F_z=-0.13$). 

While our
results for the MOs $1b_1$ and $3a_1$ also have a negative shift
the $1b_2$ and $2a_1$ orbitals is responsible for our overall 
shift not yet showing a 
complete turn-over.  This difference may come from the 
failure of the sum of eigenvalues properly representing
the total energy (cf. Eq.~\ref{eq:Ehf}).
Another explanation could be
the use of a frozen target potential in the present work.

For positive values of $F_z$
the model potential calculation also exaggerates the shift in comparison
to HF theory. Interestingly, adding correlation changes the results
by not too much, i.e., CCSD(T) follows the HF shifts at positive
values of $F_z$, and shows a greater difference in shifts on the other side. 

The net ionization rate in the right panel shows a reasonable
comparison between the methods, and for our calculations it
is dominated by the $3a_1$ orbital contribution (which is shown
for comparison without the factor of two for spin degeneracy, i.e., it is the lone orbital width). For $F_z>0$
the present net ionization rate is above the HF and CCSD(T) data
by a factor of 3-4 at the highest field strengths, since single 
ionization from the $3a_1$ orbital already exceeds those values.
 For $F_z<0$, on the other hand, the present net rate agrees
 with the HF results at moderate to strong field strengths. Starting at $-0.06$ a.u. moving towards the left, the percentage errors for our net widths with Jagau's HF values as a reference are $47 \%$, $23 \%$, $4 \%$, $3.7 \%$, and $7.3 \%$. 

One may be concerned with the disagreement between the present results
and the HF results at weak to moderate field strengths. In this regime
self-consistency effects should be weak, i.e., calculations with frozen
versus self-consistent potential ought to give similar results.
On the other hand this is the tunneling regime, and it is bound to be
very sensitive to details. 

The data for the figures are provided in Table~\ref{tab:totE} in the Appendix (Sect.\ref{sec:HFcomp}), which allows
one to disentangle the different orbital contributions.

\section{Conclusions}\label{sec:conclusions}

We used a model potential for the water molecule developed by the Madrid group~\cite{ERREA201517}, (and used 
in collaboration with the Toronto group for collision physics), to analyze the molecular properties
when exposed to a DC electric field. The calculated DC shifts and decay rates were analyzed both at the
molecular orbital level, and at the level of the total energy for net ionization. The former extends
previous model work, while the latter allows for comparison with quantum chemistry calculations~\cite{Jagau2018}.

The computations were performed using an expansion of the molecular orbitals in spherical harmonics.
For the current work we included channels up to $\ell_{\max}=3$, and we limited ourselves to only two field directions.
Future work should focus on a convergence study (higher values of $\ell_{\max}$), as well as an extension to other field directions for which 
there will be more mixing between the molecular orbitals, since for the current choice the $n a_1$ orbitals do
not mix with the $1b_{1,2}$ symmetries.

The work has also demonstrated the power of the ECS technique implemented via a finite element method.
This technology can be applied now to other small molecules of interest.

Future work should also address to what extent the DC field limit makes predictions which can be carried over to
water molecules exposed to intense infrared radiation. This can be handled using the Floquet approach, which was
applied to the case of the hydrogen molecular ion in this context~\cite{Tsogbayar_2014}.

Finally, we would like to encourage the density functional theory
community to go beyond the calculation of ground-state systems,
and to address the Stark resonance problem. A first important
step in this direction would be the extension of the present work
to the OPM approach, which should be within reach using an approximation
scheme that works well for ground-state systems~\cite{PhysRevA.62.042502},
and which avoids the need to solve an integral equation for the local potential.

\clearpage
\section{Appendix}\label{sec:app}
\subsection{Details of the matrix elements}
\label{sec:ME}
Since we implement and discuss the results of the FEM-ECS method, the main computables are the separable matrix elements of the various terms of the Hamiltonian. The oxygen potential, centrifugal potential, and kinetic energy are calculated in a straight-forward manner (however, the kinetic energy is calculated in a symmetric form derived from integration by parts). 
The main matrix element of concern is the form of the individual hydrogens, which have been expanded. For each hydrogen labelled by $j$,
\begin{equation}\label{eq:sumint}
H_j^{inlm,i'n'l'm'}=4\pi \int dr f_{in}(r)f_{i'n'}(r)\sum_{\lambda=0}^{\lambda_{\max}}\sum_{\mu=-\lambda}^{\lambda}C^{lml'm'}_{\lambda\mu}\bar{Y}_{\lambda}^{\mu}(\theta',\phi_j)\frac{V_{\lambda}(r)}{2\lambda+1},
\end{equation}
where the $f_{in}$ functions are local radial basis functions, $C$ are the results of the spherical parts of the matrix element (Gaunt integrals), the $\bar{Y}_{\lambda}^{\mu}(\theta',\phi_j)$ locate the hydrogens (by their $\theta$ and $\phi$ locations) and the $V_{\lambda}(r)$ are the radial channel dependent parts of the hydrogen potential. Our approach is to calculate the radial part of the matrix element (the integral) using a very fast parallelized 1D integrator that produces a submatrix independent of the spherical basis functions. The sums over $\lambda$ and $\mu$ values are computed after, and the full matrix is produced. The Gaunt integrals are dealt with using Wigner $3j$ coefficients computed beforehand using a code developed by Stone and Wood \cite{stone80}. The source code is provided on Wood's personal web page.\footnote{http://www-stone.ch.cam.ac.uk/wigner.shtml}
This requires a re-write of the above as,
\begin{equation}\label{eq:sumint2}
H_j^{inlm,i'n'l'm'}=\sum_{\lambda=0}^{\lambda_{\max}}\sum_{\mu=-\lambda}^{\lambda}\frac{1}{2\lambda+1}C^{lml'm'}_{\lambda\mu}\bar{Y}_{\lambda}^{\mu}(\theta',\phi_j)\cdot 4\pi \int dr f_{in}(r)V_{\lambda}(r)f_{i'n'}(r).
\end{equation}

Perhaps more clearly, a submatrix of lower structure denoted by $a$ and $b$ indices which correspond to $f_{in}$ and $f_{i'n'}$ is formed, then combined with the angular basis functions by considering a higher structure denoted by $c$ and $d$ indices corresponding $lm$ and $l'm'$. Each $lm/l'm'$ block of the higher structure contains every part of the lower structure. The full matrix is of course 2D, denoted by $\alpha$ and $\beta$ indices, but contains 4 dimensions of structure. 

The full matrix for this problem that then looks like
\begin{equation}
H_{\alpha\beta}= \left(\begin{tabular}{c c c c}
$H^{0000}$ &$H^{001-1}$&$H^{0010}$&\dots\\
$H^{1-100}$ &$H^{1-11-1}$&$H^{1-110}$&\dots\\
$H^{1000}$ &$H^{101-1}$&$H^{1010}$&\dots\\
\vdots & \vdots & \vdots &$\ddots$\\
\end{tabular}\right),
\end{equation}
where $H^{lml'm'}$ is a submatrix where $l,m,l'm'$ are fixed and $i,n,i',n'$ are not. 

The submatrices at low order have a radial part that looks like,
 \begin{small}\setlength{\tabcolsep}{1pt}
\begin{equation}\label{referencemat}
H_{k} = \left(\begin{tabular}{c c c c c}
$\langle f_{11}|H_k|f_{11}\rangle$&$\langle f_{11}|H_k|f_{13}\rangle$&$\langle f_{11}|H_k|f_{12}\rangle$&$0$&$0$\\
$\langle f_{13}|H_k|f_{11}\rangle$&$\langle f_{13}|H_k|f_{13}\rangle$&$\langle f_{13}|H_k|f_{12}\rangle$&$0$&$0$\\
$\langle f_{12}|H_k|f_{11}\rangle$&$\langle f_{12}|H_k|f_{13}\rangle$&$\langle f_{12}|H_k|f_{12}\rangle+\langle f_{21}|H_k|f_{21}\rangle$&$\langle f_{21}|H_k|f_{23}\rangle$&$\langle f_{21}|H_k|f_{22}\rangle$\\
$0$&$0$&$\langle f_{23}|H_k|f_{21}\rangle$&$\langle f_{23}|H_k|f_{23}\rangle$&$\langle f_{23}|H_k|f_{22}\rangle$\\
$0$&$0$&$\langle f_{22}|H_k|f_{21}\rangle$&$\langle f_{22}|H_k|f_{23}\rangle$&$\langle f_{22}|H_k|f_{22}\rangle$\\
\end{tabular}\right).
\end{equation}
\end{small}

The overlap between the two blocks denotes a continuity condition at the boundary between the two radial intervals (denoted by the two blocks). The letter $k$ denotes the appropriate radial Hamiltonian term, two of which are the $V_{\lambda}(r)$. The others are the centrifugal potential, the oxygen potential, and the kinetic energy. Each of these radial matrices is combined with the appropriate angular terms which involve Gaunt integrals. Thus $H^{lml'm'}$ is really an element by element sum over submatrices that look like this, or $H^{lml'm'} = \sum_k A_{k}^{lml'm'} H_{k}$, where $A_{k}^{lml'm'}$ is the appropriate angular piece. For the hydrogens $A_{k}^{lml'm'}$ contains the double sum in Eq. (\ref{eq:sumint}). The sum denoted by the capital sigma denotes the sum over every Hamiltonian term listed by $k$. $A_{k}^{lml'm'}$ is basically just the spherical harmonic overlap matrix element for all other Hamiltonian terms save for the DC potential, which contains a third spherical harmonic, i.e., $\rm \cos(\theta)$.

Two caveats of using Wigner coefficients arise. The first is that not all spherical integrals involve three spherical harmonics (specifically those involving Hamiltonian terms which lack spherical dependence). This is of course solved by using the $l=0$ spherical harmonic as the middle term in the integral (where the left and right terms are the spherical basis functions). However, this spherical harmonic is a constant, so the Wigner result must be multiplied by the inverse of the constant. A similar constant is required for the $\rm cos (\theta)$ term in the DC potential. The second caveat is that the Wigner coefficients are defined without a bar or star on the left spherical harmonic. Therefore, the appropriate response to the barring process in forming the matrix element is to multiply the $m$ quantum number of the left spherical harmonic in the Wigner coefficient by $-1$, and to multiply the entire result by $(-1)^m$, where the $m$ value corresponds to the unbarred spherical harmonic value. This arises from the standard definition used for spherical harmonics, $\bar{Y}_{l}^{m} = (-1)^m Y_{l}^{-m}$.
\subsection{Resonance positions and resonance half-widths (tables)}
Below, $-\mathfrak{Im}=\Gamma/2$, or half the resonance width.
Numbers with less than 4 significant digits have trailing zeroes. This number of sig. digs. was chosen in accord with the agreement for the real part of the energy between ECS/CAP for the orbital/field strength in Fig. \ref{fig:cap}. This applies for the real part $\mathfrak{Re}$ in the data below. In the same figure the imaginary part shows agreement for 4 decimal places with non-scientific notation (not 4 sig. digs. per the tables). The small imaginary numbers (below $10^{-7}$) should be trusted to fewer digits. 
\label{sec:tables}
\subsubsection{Positive force values}
\setlength{\tabcolsep}{6pt}
\bgroup
\def\arraystretch{0.7}
\begin{table}[!h]
\begin{center}
\begin{tabular}{l*{2}{S[round-mode=figures,round-precision=4]}} 

 MO: $1b_1$ & &\\ [0.2ex] 
\hline
\hline
   \rm $\rm F_z$  & $\mathfrak{Re}$ &$\mathfrak{Im}~$ \\ [0.2ex]
   \hline
\hline
   
    0.02 &-0.52325890277906728 & NA\\
    0.04 &-0.52565926607484736&-4.9538129827883329E-010\\
    0.06 &-0.52861114701042167&-1.3068699964488886E-006\\
    0.08&-0.53226862208335146 &-5.1104581908648882E-005\\
        0.10&-0.53680151728512493 &-3.9296599469466653E-004\\
            0.12&-0.54214534446090890 &-1.3999260518787373E-003\\
                0.14&-0.54801878240208035 &-3.3126529553757244E-003\\
                    0.16&-0.55411581227127127  &-6.1690109632551180E-003\\
         0.18&           -0.56019901072685641     & -9.9079254487538012E-003\\
0.20 &-0.56611559301139835     & -1.4402863390084524E-002\\
0.22 & -0.57177296142875822     & -1.9558063717095609E-002\\
0.24 &-0.57709859219629811     & -2.5245600477982581E-002\\
0.26  & -0.58209394655132951     & -3.1379445667165475E-002\\
0.28 &-0.58671209984239092     & -3.7899070665166207E-002\\
0.30 &-0.59095530412333930     & -4.4689854268142114E-002\\

\hline
\hline

\end{tabular}

\caption{Resonance positions and half-widths for the molecular orbital $1b_1$ of $\rm H_2O$ as a function of $F_z>0$. $\mathfrak{Re}$ refers to the real part of the energy, and $\mathfrak{Im}$ refers to the imaginary part of the energy. }
\end{center}
\end{table}
\setlength{\tabcolsep}{6pt}
\bgroup
\def\arraystretch{0.7}
\begin{table}[!h]
\begin{center}
\begin{tabular}{l*{2}{S[round-mode=figures,round-precision=4]}} 

 MO: $3a_1$ & &\\ [0.2ex] 
\hline
\hline
   \rm $\rm F_z$  & $\mathfrak{Re}$ &$\mathfrak{Im}~$ \\ [0.2ex]
   \hline
\hline

 0.02&-0.56806179941273316  &    -1.4363028658642402E-012\\
  0.04&-0.56511430284630382   &   -6.4736526797257262E-008\\
 0.06&  -0.56357893658174607     & -5.9492263407908406E-005\\
    0.08&-0.56467741899858281     & -1.2927831948746061E-003\\
     0.10&-0.56794362049464397     & -6.5977672437574101E-003\\
     0.12&-0.57035388810433052     & -1.7557121234624724E-002\\
     0.14&-0.56850504506604160     & -3.3690051139452393E-002\\
      0.16&-0.55901441707872268     & -5.2635289171937778E-002\\
      0.18 &-0.54000558562549617     & -6.7795421510000384E-002\\
  0.20 & -0.51779322343897471     & -7.3444656411718795E-002  \\
  0.22 & -0.49961685369122222     & -7.2862550182455632E-002\\
  0.24 & -0.48524148456520211     & -7.0441967251027818E-002\\
  0.26 & -0.47378598188926530     & -6.8044401886565090E-002\\
  0.28 &-0.46396680231589704     & -6.6215924796290807E-002\\
  0.30&-0.45531589128987804     & -6.5054799084613604E-002\\

\hline
\hline

\end{tabular}

\caption{Resonance positions and half-widths for the molecular orbital $3a_1$ of  $\rm H_2O$ as a function of $F_z>0$. $\mathfrak{Re}$ refers to the real part of the energy, and $\mathfrak{Im}$ refers to the imaginary part of the energy.}
\end{center}
\end{table}
\setlength{\tabcolsep}{6pt}
\bgroup
\def\arraystretch{0.7}
\begin{table}[!h]
\begin{center}
\begin{tabular}{l*{2}{S[round-mode=figures,round-precision=4]}} 

 MO: $1b_2$ & &\\ [0.2ex] 
\hline
\hline
   \rm $\rm F_z$  & $\mathfrak{Re}$ &$\mathfrak{Im}~$ \\ [0.2ex]
   \hline
\hline

0.02   &-0.71604709736398753     & -3.0668768012778290E-012\\
0.04 & -0.72272267363722398     & -4.7469292813543686E-012\\
0.06 &-0.72995177664043986     & -9.4594050088363522E-011\\
0.08&-0.73778204200255759     & -6.6643337177144528E-008\\
0.10&-0.74628262933897593     & -2.9078488657122290E-006\\
0.12&-0.75555764897842903     & -3.3119873520940706E-005\\
0.14&-0.76573255286071928     & -1.7772753431505542E-004\\
0.16&-0.77689795992904254     & -5.9999527008557357E-004\\
0.18&-0.78906135334748673     & -1.5004230124128696E-003\\
0.20&-0.80214888695522502     & -3.0539890558808520E-003\\
0.22&-0.81603903292845692     & -5.3860797520583073E-003\\
0.24&-0.83059528931146931     & -8.5543808815248500E-003\\
0.26&-0.84570403188356802     & -1.2579215910138990E-002\\
           0.28& -0.86125399378217360     & -1.7453128816950811E-002\\
           0.30&-0.87716901541320857    & -2.3131737126502253E-002\\
\hline
\hline

\end{tabular}

\caption{Resonance positions and half-widths for the molecular orbital $1b_2$ of $\rm H_2O$ as a function of $F_z>0$. $\mathfrak{Re}$ refers to the real part of the energy, and $\mathfrak{Im}$ refers to the imaginary part of the energy.}
\end{center}
\end{table}
\clearpage
\subsubsection{Negative force values}
\setlength{\tabcolsep}{6pt}
\bgroup
\def\arraystretch{0.7}
\begin{table}[!h]
\begin{center}
\begin{tabular}{l*{2}{S[round-mode=figures,round-precision=4]}} 

 MO: $~1b_1$ & &\\ [0.2ex] 
\hline
\hline
   $\rm F_z$ & $\mathfrak{Re}$ &$\mathfrak{Im}$ \\ [0.2ex]
   \hline
\hline
   
    -0.02 &-0.51987145873888529 &-2.7691682230616999E-012\\
    -0.04 &-0.51884592107248872 &-3.9402878659001128E-010\\
    -0.06 &-0.51828569676469372 &-9.5114450487585490E-007\\
    -0.08&-0.51826726514153576 &-3.4889302798618303E-005\\
        -0.10&-0.51886669063033786 &-2.5463078950368388E-004\\
            -0.12&-0.52001220429113815 &-8.7111235919475137E-004\\
               -0.14&-0.52152109260346358 &-1.9937539212812627E-003\\
                    -0.16&-0.52321527828064462  &-3.6230012223050015E-003\\
         -0.18&           -0.52497691813345104     & -5.6989781471608074E-003\\
-0.20 &-0.52672561117307704     & -8.1566135210368144E-003\\
-0.22 &-0.52842680550187404     & -1.0922313962977589E-002\\
-0.24 &-0.53005730790360450    & -1.3953447308526178E-002\\
-0.26  &-0.53160185949437133     & -1.7191939391923108E-002\\
-0.28 &-0.53307324443208837     & -2.0607579687673259E-002\\
-0.30 &-0.53445759106338553     & -2.4180361504564339E-002\\

\hline
\hline

\end{tabular}

\caption{Resonance positions and half-widths for the molecular orbital $1b_1$ of  $\rm H_2O$ as a function of $F_z<0$. $\mathfrak{Re}$ refers to the real part of the energy, and $\mathfrak{Im}$ refers to the imaginary part of the energy.}
\end{center}
\end{table}

\setlength{\tabcolsep}{6pt}
\bgroup
\def\arraystretch{0.7}
\begin{table}[!h]
\begin{center}
\begin{tabular}{l*{2}{S[round-mode=figures,round-precision=4]}} 

 MO: $~3a_1$ & &\\ [0.2ex] 
\hline
\hline
   $\rm F_z$ & $\mathfrak{Re}$ &$\mathfrak{Im}$ \\ [0.2ex]
   \hline
\hline
   
    -0.02 &-0.57618081202548399&-2.4921891620345441E-013\\
    -0.04 &-0.58105466794254612&-1.4370509122141963E-008\\
    -0.06 &-0.58652249021287906&-1.3352679464726484E-005\\
    -0.08&-0.59287262238114835&-3.0373596430698769E-004\\
        -0.10&-0.60005818362209851&-1.6429205026623746E-003\\
            -0.12&-0.60733457926631162&-4.5535045879802226E-003\\
               -0.14&-0.61390433843106229 &-8.9281319297057730E-003\\
                    -0.16&-0.61928877369488089  &-1.4387227882989359E-002\\
         -0.18&-0.62328112410249614& -2.0481117961812852E-002\\
-0.20 &-0.62581001741699749& -2.6871012225925840E-002\\
-0.22 &-0.62695985451198732&   -3.3217895929690697E-002\\
-0.24 &-0.62679922137171362&-3.9348436221515681E-002\\
-0.26  &-0.62549514854363775&-4.5027544537835022E-002\\
-0.28 &-0.62320059889137358&-5.0234123314665341E-002\\
-0.30 &-0.62002693790660957&-5.4831293900205309E-002\\

\hline
\hline

\end{tabular}

\caption{Resonance positions and half-widths for the molecular orbital $3a_1$ of  $\rm H_2O$ as a function of $F_z<0$. $\mathfrak{Re}$ refers to the real part of the energy, and $\mathfrak{Im}$ refers to the imaginary part of the energy.}
\end{center}
\end{table}
\clearpage
\setlength{\tabcolsep}{6pt}
\bgroup
\def\arraystretch{0.7}
\begin{table}[!h]
\begin{center}
\begin{tabular}{l*{2}{S[round-mode=figures,round-precision=4]}} 

 MO: $~1b_2$ & &\\ [0.2ex] 
\hline
\hline
   $\rm F_z$ & $\mathfrak{Re}$ &$\mathfrak{Im}$ \\ [0.2ex]
   \hline
\hline
   
    -0.02 &-0.70422392105131359 &$NA$\\
    -0.04 &-0.69903041686089873&-4.3531859311771776E-012\\
    -0.06 &-0.69429694286376176 &-1.1164658660228287E-010\\
    -0.08&-0.69001961721170091 &-6.5539032384771293E-008\\
        -0.10&-0.68620790458193925 &-2.5266887401219614E-006\\
            -0.12&-0.68288954751486097 &-2.5668694284075424E-005\\
               -0.14&-0.68009420960270206 &-1.2403351354929449E-004\\
                    -0.16&-0.67782007123468735  &-3.8223559633775530E-004\\
         -0.18&           -0.67602155056267543     & -8.8416997531863626E-004\\
-0.20 &-0.67462368446167320     & -1.6855844540283342E-003\\
-0.22 &-0.67354419171392310     & -2.8132403980627159E-003\\
-0.24 &-0.67270678940423811    & -4.2641939101489479E-003\\
-0.26  &-0.67205162195989099    & -6.0253251093710320E-003\\
-0.28&-0.67152528441428216     & -8.0714868004022385E-003\\
-0.30 &-0.67109598581188823     & -1.0373476888983846E-002\\

\hline
\hline

\end{tabular}

\caption{Resonance positions and half-widths for the molecular orbital $1b_2$ of  $\rm H_2O$ as a function of $F_z<0$. $\mathfrak{Re}$ refers to the real part of the energy, and $\mathfrak{Im}$ refers to the imaginary part of the energy.}
\end{center}
\end{table}

\subsection{Comparison with HF total ionization rates}
\label{sec:HFcomp}

Table~\ref{tab:totE} contains the Stark shifts and widths of relevant orbitals, and total Stark shifts and resonance widths, to be compared with with the HF results of Jagau \cite{Jagau2018}. We carry out a direct sum (written $\rm DS ~(2)$ for a sum of all orbitals, where the two indicates double occupation due to spin degeneracy) of occupied orbitals to find the net energy and shifts. The bottom two orbitals ($1a_1$ and $2a_1$) are left out when their values are too small to be comparable to the net HF values, which only happens for the widths and not the Stark shifts. The HF row references Jagau \cite{Jagau2018}. 
\clearpage
\setlength{\tabcolsep}{4pt}
\bgroup
\def\arraystretch{0.7}
\begin{table}[!h]
\begin{center}

 \begin{tabular*} {\columnwidth}{@{\extracolsep{\fill}\extracolsep{0pt}}llllllll}

\hline

 &&$\mathfrak{Re}~\Delta \rm E$&\\ [0.2ex] 
\hline

  $F_z$&&0.04 &  0.06& 0.08&  0.10& 0.12 &0.14\\ [0.2ex]
   \hline

   $1a_1$ &&$-0.000013$&$-0.000021$&$-0.000028$&$-0.000037$&$-0.000045$&$-0.000055$\\
    $2a_1$ &&$-0.016565$&$-0.025636$&$-0.035222$&$-0.045316$&$-0.055919$&$-0.067036$\\
    $1b_2$ & &$-0.012833$&$-0.020062$&$-0.027892$&$-0.036393$&$-0.045668$&$-0.055843$\\
    $3a_1$ &&$0.006711$&$0.008246$&$0.007148$&$0.003881$&$0.001471$&$0.003320$\\
    $1b_1$&&$-0.004319$&$-0.007271$&$-0.010928$&$-0.015461$&$-0.020805$&$-0.026680$\\
        
         $\rm DS$ (2)&&$-0.054039$&$-0.089488$&$-0.133846$&$-0.186651$&$-0.241933$&$-0.292585$\\
            HF &&$-0.038045$&$-0.062476$&$-0.091129$&$-0.124048$&$-0.160648$&$-0.200088$\\

\hline

 &&$\Gamma$\\ [0.2ex] 
\hline

  $F_z$& &0.04 &  0.06& 0.08&  0.10& 0.12 &0.14\\ [0.2ex]
   \hline

   %$1a_1$ &&0.000000&0.000000&0.000000&0.000000&0.000000&0.000000\\
    %$2a_1$ &&0.000000&0.000000&0.000000&0.000000&0.000000&0.000000\\
    $1b_2$ &&0.000000&0.000000&0.000000&0.000006&0.000066&0.000355\\
    $3a_1$& &0.000000&0.000119&0.002586&0.013196&0.035114&0.067380\\
    $1b_1$&&0.000000&0.000003&0.000102&0.000786&0.002800&0.006625\\
     $\rm DS ~ (2)$&&0.000000&0.000243&0.005376&0.027975&0.075961&0.148723\\
      
            HF &&0.000000&0.000040&0.001021&0.005029&0.013694&0.027745\\

\hline

 &&$\mathfrak{Re}~\Delta \rm E$&\\ [0.2ex] 
\hline

  $F_z$& &$-0.04 $&$  -0.06$&$ -0.08$&$  -0.10$&$ -0.12 $&$-0.14$\\ [0.2ex]
   \hline

   $1a_1$ &&0.000011&0.000016&0.000020&0.000024&0.000028&0.000031\\
    $2a_1$ &&0.014373&0.020694&0.026410&0.031497&0.035926&0.039666\\
    $1b_2$ & &0.010859&0.015593&0.019870&0.023682&0.027000&0.029797\\
    $3a_1$ &&$-0.009230$&$-0.014697$&$-0.021048$&$-0.028233$&$-0.035510$&$-0.042079$\\
    $1b_1$&&0.002494&0.003054&0.003073&0.002473&0.001328&$-0.000181$\\
        $\rm DS$ (2)&&0.037016&0.049319&0.056651&0.058886&0.057544&0.054463\\

            HF &&0.024155&0.030418&0.031849&0.028404&0.021187&0.011549\\

\hline

 &&$\Gamma$\\ [0.2ex] 
\hline

  $F_z$& &$-0.04 $&$  -0.06$&$ -0.08$&$  -0.10$&$ -0.12 $&$-0.14$\\ [0.2ex]
   \hline

   %$1a_1$ &&0.000000&0.000000&0.000000&0.000000&0.000000&0.000000\\
    %$2a_1$ &&0.000000&0.000000&0.000000&0.000000&0.000000&0.000000\\
    $1b_2$ &&0.000000&0.000000&0.000000&0.000005&0.000051&0.000248\\
    $3a_1$&&0.000000&0.000027&0.000607&0.003286&0.009107&0.017856\\
    $1b_1$&&0.000000&0.000002&0.000070&0.000509&0.001742&0.003988\\

             $\rm DS~ (2)$&&0.000000&0.000057&0.001355&0.007600&0.021801&0.044185\\
            HF &&0.000006&0.000108&0.001759&0.007911&0.021029&0.041178\\

\hline

\end{tabular*}

\caption{\label{tab:totE}
Stark shifts and widths for the valence orbitals of the water molecule for different force strengths, along with the total molecular Stark shifts and widths under DS (2). All results are from the current work except for HF which references Jagau \cite{Jagau2018}.}
\end{center}
\end{table}

\clearpage
\begin{acknowledgments}
Discussions with Tom Kirchner and Michael Haslam are gratefully acknowledged. We would also like to thank Steven Chen for support with the high performance computing server used for our calculations.
Financial support from the Natural Sciences and Engineering Research Council of Canada is gratefully acknowledged. \end{acknowledgments}

%
% BibTeX users please use
%\bibliographystyle{plain}
%\bibliographystyle{unsrt}

\bibliography{PartialWaveMethod}

\end{document}